\documentclass[journal]{IEEEtran}

\usepackage{graphicx}
\usepackage{placeins}
\usepackage{amsmath}
\usepackage[linesnumbered,ruled]{algorithm2e}
\usepackage{booktabs}

\ifCLASSINFOpdf
\else
\fi

\begin{document}
\title{Modeling of Negative Capacitance of Ferroelectric Capacitors as a Non-Quasi Static Effect}

\author{Borna~Obradovic,
        Titash~Rakshit,
				Ryan~Hatcher,
				Jorge~Kittl,
        and~Mark~S.~Rodder%} <-this % stops a space
\thanks{B. Obradovic, T. Rakshit, R. Hatcher, J. Kitttl and M. S. Rodder are with the 
Samsung Advanced Logic Lab, Austin TX.}% <-this % stops a space
%\thanks{D. Lin, N. Waldron and N. Collaert are with, IMEC Leuven, Belgium.}% <-this % stops a space
}
\maketitle

\begin{abstract}
Pulse-based studies of ferroelectric capacitor systems have been used by several groups to experimentally
probe the mechanisms of apparent negative capacitance. In this paper, the behavior of such systems
is modeled through SPICE simulation with a delayed-response Preisach model, and the results are compared to available data. 
It is found that a simple ferroelectric
domain delay model can explain much of the observed behavior, capturing the qualitatively different effects of bipolar
and unipolar switching, as well as the voltage dependence of said switching. The observed behavior and its
modeling suggests that the observed negative capacitance is in fact associated with ferroelectric switching,
and its presence is highly sensitive to the switching frequency and other details of the ferroelectric system.

\end{abstract}


% \fi
\IEEEpeerreviewmaketitle

\section{Introduction}
\label{Intro}
\begin{figure}[!ht]
\centering
\includegraphics[width=2.5in]{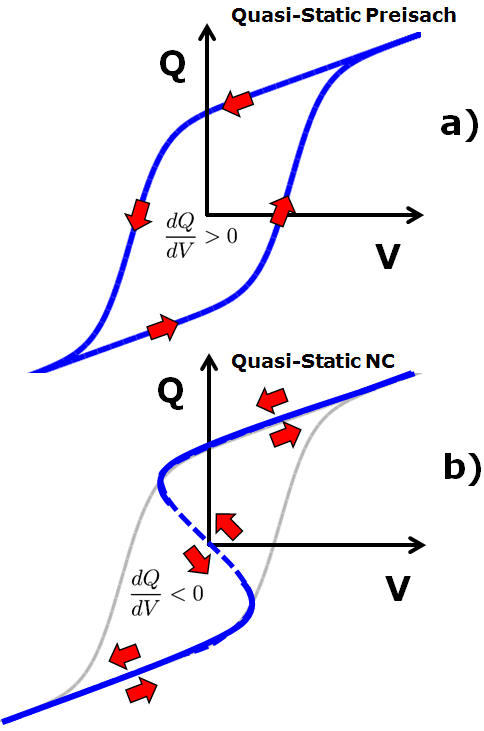}
% where an .eps filename suffix will be assumed under latex, 
% and a .pdf suffix will be assumed for pdflatex; or what has been declared
% via \DeclareGraphicsExtensions.
\caption{The quasi-static Q-V curves for FeCaps are shown, based on the Preisach model (sub-plot a),
and the stabilized Negative Capacitance model (sub-plot b). Saturation loops are shown. In the Preisach
model, traversal of the saturation loop is always in the indicated direction; reversing the traversal direction
prior to full saturation creates minor loops. Quasi-static capacitance in the Preisach model  is always positive. 
In the stabilized NC model,
a non-hysteretic path is available. Traversal of the path is bi-directional. Quasi-static capacitance is negative
along the dashed portion of the path. }
\label{PreisachModel}
\end{figure}

\IEEEPARstart{N}{egative} capacitance has been postulated theoretically \cite{Datta} and extensively studied experimentally
(\cite{Sayeef}, \cite{ferro1}, \cite{SDattaEDL}).
The theoretical basis for Negative Capacitance (NC) is an assumed Energy-Charge (U-Q) relationship
with a region of negative curvature (U-Q ansatz) \cite{Datta}. As a consequence of the negative curvature,
a switching path with negative capacitance (Fig. \ref{PreisachModel}b) becomes available. This is in contrast to the
standard Preisach model of a ferroelectric capacitor (Fig. \ref{PreisachModel}a). However, this path
is unstable and not directly observable in stand-alone FeCaps. Arguments have been made 
\cite{Datta}  that
this unstable path can be stabilized by connecting the FeCap in series with a standard (positive U-Q curvature)
capacitor which satisfies specific matching criteria \cite{Stabilization}, 
thereby making the negative capacitance branch traversable. As a consequence, using such a capacitor
arrangement in the gate stack of a FET would result in increased stack capacitance and sub-60 mV/dec 
subthreshold slope (\cite{Datta}, \cite{Sayeef}). However, precisely how the stabilized state is established at the microscopic
level is unclear and challenging to explain. In the face of this difficulty, 
the goal of this paper is to investigate whether
some of the key experimental findings can in fact be reproduced without the U-Q ansatz, consequently doing away
with the need for a microscopic model of the stabilized state. 
The alternative conceptual model proposed
herein is simply a Preisach Ferroelectric (\cite{Preisach}, \cite{Preisach2}) in which domain switching takes place with a delay.
The existence of a switching delay of ferroelectric domains is itself well known and characterized, with little ambiguity with regards to the microscopic model (\cite{ferro2}, \cite{ferro3}, \cite{Namlab}).

\begin{figure}[!ht]
\centering
\includegraphics[width=3.5in]{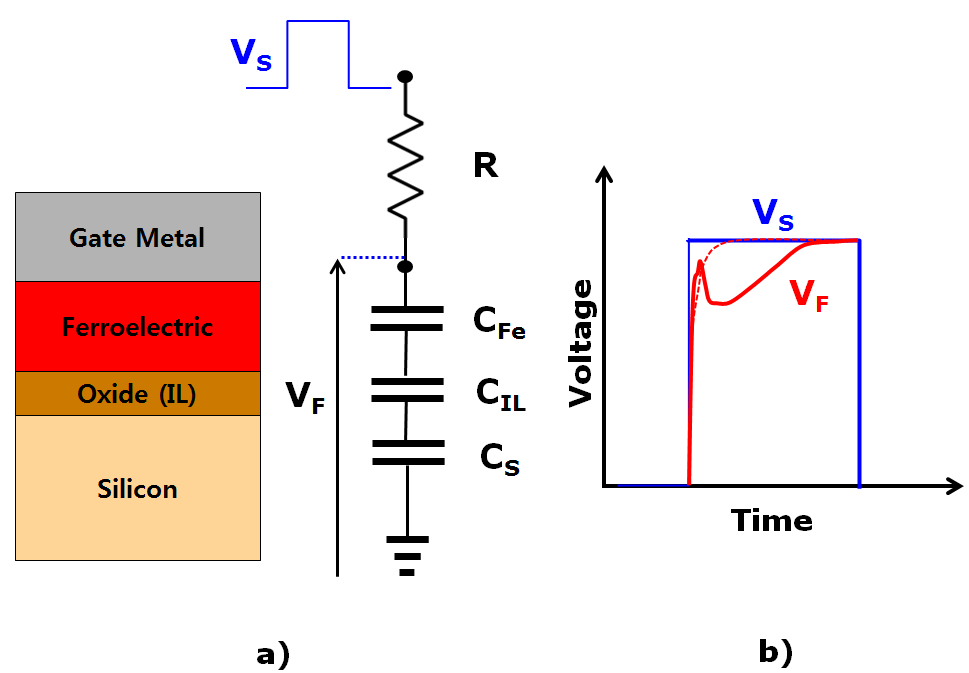}
% where an .eps filename suffix will be assumed under latex, 
% and a .pdf suffix will be assumed for pdflatex; or what has been declared
% via \DeclareGraphicsExtensions.
\caption{The experimental setup for transient pulse experiments is illustrated schematically. The capacitor stack
consists of a ferroelectric layer (HfZrO in present case), a non-ferroelectric dielectric layer (IL), and a
partially depleted Si substrate. Voltage pulses are applied to the stack through an access resistor while the stack 
voltage (V$_{F}$ is monitored. The expected V$_F$ response is shown with the dashed line; the actual
response is (qualitatively) shown with the solid red line. A ``spike", followed by a dip and a recovery, is observed.}
\label{Experiment}
\end{figure}

While experimental studies range from basic FeCap measurements to full CMOS Ring Oscillator circuits  \cite{GF}, 
a particularly
data-rich set of experiments have been performed by studying the transient response of FeCap stacks to pulse
waveforms (\cite{Sayeef}, \cite{SDattaEDL}). 
The basic experimental setup for such measurements is illustrated schematically in Fig. \ref{Experiment}a.
A stack of dielectrics is used to
form the overall capacitor, which is then pulsed using a square-wave voltage waveform of varying amplitude and frequency.
The voltage across the overall stack is observed as a function of time. 

The qualitative behavior of the measurements (valid for all groups cited in references) is illustrated
in Fig. \ref{Experiment}b. 
As noted in all the studies which perform experiments of this type, the observed voltage waveform is quite anomalous from the standpoint of a ``positive capacitance" RC  behavior.
Specifically, instead of a monotonic response, the voltage initially spikes, followed by a dip with a subsequent gradual
rise (or fall) to VDD (or -VDD) (Fig. \ref{Experiment}b). This ``spike" behavior has been characterized as ``inductive-like" \cite{Sayeef}, 
and NC has been used
to explain it. In this paper, the anomalous behavior is investigated using a delayed Preisach model instead.

Experiments of this sort have been performed by several groups, but the recent measurements of \cite{SDattaEDL} are particularly helpful due to their more comprehensive
parametric sweeps. Specifically, the measurements in \cite{SDattaEDL} are performed for both bipolar and unipolar switching,
across a range of pulse voltages. The data from \cite{SDattaEDL}  are therefore used as the primary source of for model calibration
and evaluation.

\FloatBarrier
\section{Model Description}
\label{Modeling}

\begin{figure}[!ht]
\centering
\includegraphics[width=2.5in]{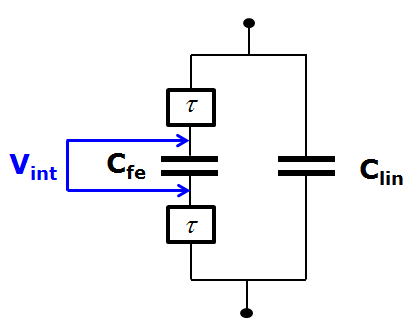}
% where an .eps filename suffix will be assumed under latex, 
% and a .pdf suffix will be assumed for pdflatex; or what has been declared
% via \DeclareGraphicsExtensions.
\caption{The conceptual model of the ferroelectric capacitor is illustrated. The model consists of two
parts: the ferroelectric polarization component (C$_{fe}$), and the non-ferroelectric component (C$_{lin}$).
Delay elements (shown as $\tau$) are used to introduce a delay in the ferroelectric response. The 
instantaneous voltage applied to the quasistatic ferroelectric model is labeled V$_{int}$ (for ``internal"). 
Note that the label internal does not refer to a spatial location; it is only internal in the conceptual model.  }
\label{Model_Concept}
\end{figure}

The overall conceptual model for the FeCap consists of two components: a delayed ferroelectric polarization capacitance, and
a quasistatic non-ferroelectric capacitance (as shown in Fig. \ref{Model_Concept}). The ferroelectric response
is delayed due to the intrinsic switching dynamics of the Fe domains, while the non-ferroelectric component is
governed by fast electronic polarization. The delayed ferroelectric response itself is modeled as a 
quasistatic Preisach FeCap combined with an explicit delay. The approach is somewhat different than that of recent 
work on introducing RC delay, such as that of \cite{alternative}. In the latter approach, an explicitly time-varying 
series resistor is added to the FeCap (along with an explicitly time-varying FeCap). Other than the practical limitation
of explicit time dependence for model elements, such models attribute time-dependent behavior of FeCaps
to physical resistance components (although it is also possible to interpret the added resistance simply as a means
of introducing a first-order delay to a SPICE model, without a literal physical resistance). In this work, the delay is assumed to be due to the internal switching dynamics of 
ferroelectric domains, but no attempt is made to explain the detailed microscopic origin of the switching
delay (which is presented in \cite{Namlab}). The implementation of the model is not explicitly time-dependent, so it is applicable under arbitrary biasing conditions. Furthermore, the hysteretic nature of the model enables the exploration of history-dependent
effects, such as the transition from bipolar to unipolar switching.

\begin{figure}[!ht]
\centering
\includegraphics[width=3.0in]{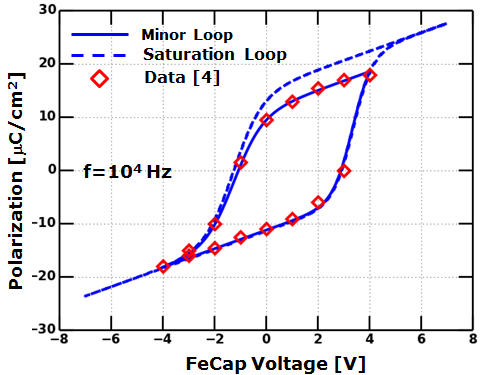}
% where an .eps filename suffix will be assumed under latex, 
% and a .pdf suffix will be assumed for pdflatex; or what has been declared
% via \DeclareGraphicsExtensions.
\caption{The measured and simulated P-V characteristics of the HfZrO FeCap are compared. The data
(shown as red diamonds) is sampled from \cite{SDattaEDL}. The data does not appear to be on the saturation
loop. Saturation loop model parameters are estimated so that the minor loop simulation using the Preisach model
(in the interval V $\epsilon$ [-4V, 4V]) best fits the data. Saturation and minor loop simulation is shown with
blue lines.}
\label{PVModel}
\end{figure}

\begin{figure}[!ht]
\centering
\includegraphics[width=3.2in]{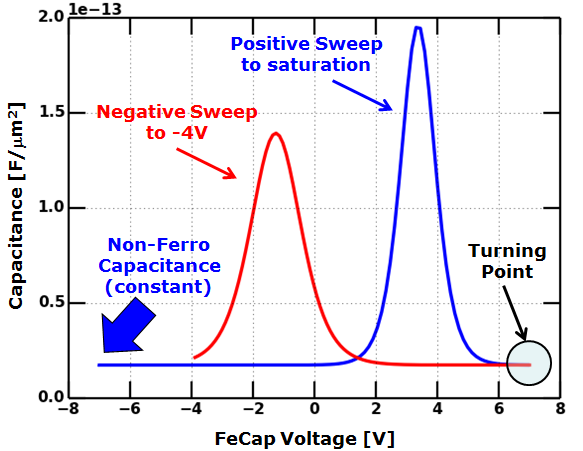}
% where an .eps filename suffix will be assumed under latex, 
% and a .pdf suffix will be assumed for pdflatex; or what has been declared
% via \DeclareGraphicsExtensions.
\caption{The simulated quasistatic capacitance characteristics of the  HfZrO FeCap are illustrated, based on the 
calibration of Fig. \ref{PVModel}. Both the linear and Ferroelectric polarizations are included. The quasistatic
capacitance is always positive in the model.}
\label{Capacitance}
\end{figure}

The quasi-static ferroelectric polarization
is described next. The ``raw" response function is given by:
\begin{equation}
F^{\pm}(V_{int})= \theta^\pm \cdot tanh \bigg( \frac{V_{int} \mp V_c^\pm}{V_{sc}^\pm} \bigg)
\label{raw}
\end{equation}

\noindent where $V_{int}$ is the voltage representing the internal state of the FeCap (related
to the applied voltage, as described next), and $V_c^\pm$ and $V_{sc}^\pm$ are model parameters describing
the coercive voltages and the voltage scales, respectively. Likewise, $\theta^\pm$ is a model parameter
which sets the polarization strength in each state. Each quantity in Eqn. \ref{raw}
has a ``plus" and ``minus" label, depending on whether the capacitor last experienced an increase
or decrease in applied voltage (respectively). This is referred to as the ``state" of the FeCap.
The actual ferroelectric polarization $P_{FE}$ is computed using Eqn. \ref{Polarization}:

\begin{equation}
P_{FE}(V_{int}) = \bigg( F^{\pm}(V_{int})-F^+_j \bigg) \cdot 
\bigg[ \frac{P_j-P_i}{F^+_j-F^-_i} \bigg] + P_j
\label{Polarization}
\end{equation}

\noindent where the indices $i$, $j$ denote the currently active turning points, with $V_j > V_i$.
Likewise, the quantities $F^+_j$, $F^-_i$, $P_j$, and $P_i$ are evaluated at the current active
pair of turning points. Eqn. \ref{Polarization} simply provides scaling and shifting of the raw response
of Eqn. \ref{raw} to ensure that the polarization curve passes through the active pair of turning points
(this is necessary to provide reasonable minor-loop behavior). 
As is standard for turning-point models, turning points are created and destroyed
dynamically as the internal voltage of the FeCap switches. The ferroelectric response is thus modeled
as a quasi-static function of the ``internal" voltage of the FeCap (described by $V_{int}$). The overall
behavior of the FeCap is therefore dependent on the dynamics of $V_{int}$. 
The quasistatic parameters of
the FeCap model are calibrated using the data of \cite{SDattaEDL},
as shown in Fig. \ref{PVModel}. The associated capacitance (likewise state-dependent) is shown in Fig. 
\ref{Capacitance}.

In this work, $V_{int}$ is
governed by a second-order delay of the applied voltage across the FeCap. This is simply an empirical dynamical
system, establishing the relationship between the ``internal" and applied voltages across the FeCap. A $2^{nd}$
order system is chosen to provide a more explicit delay, as opposed to $1^{st}$ order ``damping". However, this
should not be interpreted as arising from microscopic physical considerations; a detailed, physically-based 
voltage-dependent delay model is yet to be developed.
The differential
equation relating the ``internal" voltage $V_{int}$ and the applied voltage $V_{app}$ is given as:

\begin{equation}
\ddot{V}_{int} 
+  2 \gamma \omega_0 \dot{V}_{int} 
+ \omega_0^2 V_{int}
= \omega_0^2 V_{app}
\label{dynamics}
\end{equation}

\noindent where the natural frequency $\omega_0$ and the damping ratio $\gamma$ are calibration parameters
of the model. The ``memory" property of the FeCap is handled by the model though the turning point
history and the active state. Both are changed in a discrete manner, when the temporal derivative of
$V_{int}$ changes sign. This ensures that the state itself is not experiencing the second-order dynamics
described in Eqn. \ref{dynamics}; the dynamics merely provide a delay for an abrupt switching of the state.
Finally, the total
charge of the capacitor is computed as:
\begin{equation}
Q_{tot}(V_{int}, V_{app}) = \bigg(P(V_{int}) + C_{lin} V_{app} \bigg) \cdot A
\label{Charge}
\end{equation}

\noindent where $C_{lin}$ is the non-ferroelectric capacitance, and $A$ is the total capacitor area. 
The non-ferroelectric response is modeled as being driven by the instantaneous applied voltage $V_{app}$,
since the non-ferroelectric response is assumed to be much faster than any modulation of the applied voltage.
For the purpose of this work, the complete model is implemented in Verilog-A and used within Synopsys HSPICE.

\FloatBarrier
\section{Results and Discussion}

The FeCap model described in Sec. \ref{Modeling} is next applied to the SPICE simulation of the experimental
setup. The cases of bipolar and unipolar switching are considered separately. The parameters which govern
model dynamics ($\omega_0$, $\gamma$) are calibrated for the best match to the overall transient data set.
The obtained values (as shown in Sec. \ref{sec:Bipolar}) are $\omega_0=12 \cdot 10^6$ rad/sec, $\gamma=3$.
The corner frequency of ~2 MHz is consistent with values reported in the literature for HfZrO and HfO systems (\cite{ferro1}, \cite{Namlab}).
Strong damping is needed to prevent unphysical ``ringing". The full set of model parameters is summarized
in Table \ref{ParamTable}. The initial fitting is performed against the P-V data (Fig. \ref{PVModel}). 
Parameters which govern switching dynamics ($\gamma$, $\omega_0$) are obtained purely
from pulse data. 

\begin{table}[htbp]
  \centering
%  \caption{Add caption}
    \begin{tabular}{cccc}
    \toprule
    \textbf{Parameter} & \textbf{Units} & \textbf{Fit to P-V Hysteresis} & \textbf{Fit to Transient Pulse} \\
    \midrule
    $\theta^+$ & $\mu$C/cm$^2$ & 14    & 15 \\
    $\theta^-$ & $\mu$C/cm$^2$ & -10   & -10 \\
    $V_c^+$   & V     & 3.3   & 3.3 \\
    $V_c^-$   & V     & 1.2   & 2 \\
    $V_{sc}^+$  & V     & 0.75  & 0.75 \\
    $V_{sc}^-$  & V     & 0.9   & 0.9 \\
    $C_{lin}$  & fF/$\mu$m$^2$ & 17.5  & 17.5 \\
    $\gamma$ &   &  & 3 \\
    $\omega_0$ & rad/sec &   & 12 $\cdot$ 10$^6$  \\
    \bottomrule \\
    \end{tabular}%
    \caption{The model parameters extracted by fitting to the P-V and transient pulse data of \cite{SDattaEDL}
    are shown.
    }
  \label{ParamTable}%
\end{table}%

For best agreement to pulse data however, some modification of the parameters obtained from P-V
data is required. Most of the parameters show good consistency between the ``DC" fitting of the P-V
 hysteresis and transient data (Table \ref{ParamTable}). The only notable discrepancy is that of  $V_c^-.$ A possible explanation 
 for this discrepancy is the weakness of the Preisach model in ``data-free" prediction of minor loop behavior.
 Since the calibration was actually done on a minor loop, the saturation loop may be in error. Experience with fitting
 the model suggests that the real data shows a more skewed P-V curve with stronger $\theta^+$ polarization
 than is currently being modeled, requiring the use of a larger $V_c^-$ for best fit to the pulse data. 
 Further measurements and modeling will be needed to resolve this discrepancy.

%\FloatBarrier
\subsection{Bipolar Switching}
\label{sec:Bipolar}

\begin{figure}[!ht]
\centering
\includegraphics[width=3.0in]{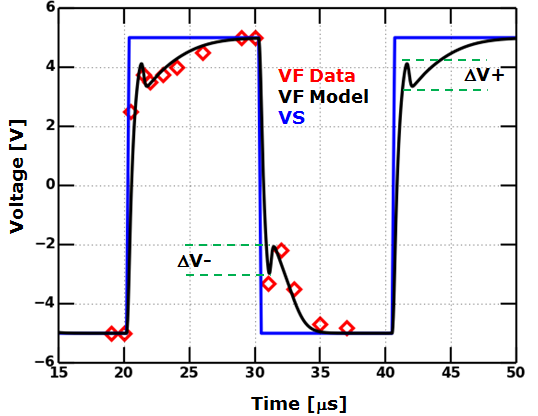}
% where an .eps filename suffix will be assumed under latex, 
% and a .pdf suffix will be assumed for pdflatex; or what has been declared
% via \DeclareGraphicsExtensions.
\caption{The measured \cite{SDattaEDL} and simulated transient response for the ferroelectric capacitor stack
to a bipolar square-pulse excitation is shown. The ``anomalous" spikes in the early part
of the response to each pulse are clearly visible in data and simulation. The spike magnitudes (for the positive
and negative pulses) are labeled $\Delta V+$ and $\Delta V-$, respectively.}
\label{Bipolar}
\end{figure}

\begin{figure}[!ht]
\centering
\includegraphics[width=3.0in]{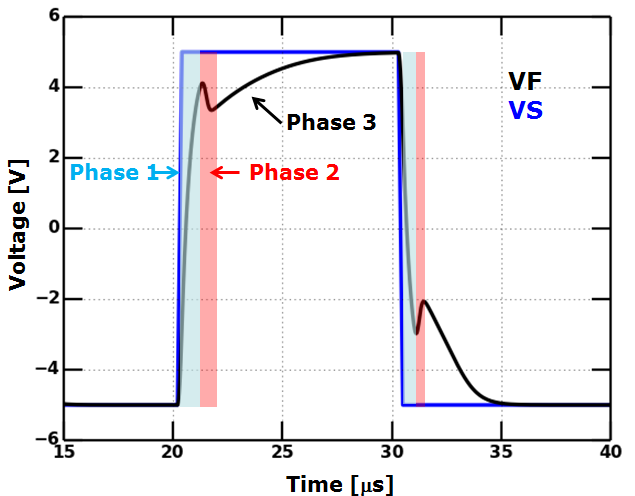}
% where an .eps filename suffix will be assumed under latex, 
% and a .pdf suffix will be assumed for pdflatex; or what has been declared
% via \DeclareGraphicsExtensions.
\caption{The phases of the transient event are illustrated. The first phase (blue shading) is the initial rise, prior
to the activation of ferroelectric domains. The second phase (red shading) is the initial ferroelectric response, during
which the ferroelectric capacitance increases quickly. The thirds phase (unshaded region) is the final ferroelectric
response, which takes place on a longer timescale on which ferroelectric domains switch nearly quasistatically.}
\label{Bipolar_Stages}
\end{figure}

For bipolar switching, the square waveform ranges from -VDD to +VDD, exercising both the positive and
negative domain switching thresholds. The sampled data and simulation results for VDD=5V are 
shown in Fig. \ref{Bipolar}. As discussed in Sec. \ref{Intro}, the measured data exhibits an initial
voltage spike, followed by a more gradual rise to the peak pulse voltage. As can be seen in Fig. \ref{Bipolar}, 
SPICE simulation indicates a very similar behavior. Arguing from the standpoint of a delayed ferroelectric
response, the ``spike" behavior can be understood in the following sequence (numbered phases illustrated in Fig. \ref{Bipolar_Stages}).

\begin{enumerate}
\item{Initial Rise. The input waveform has a rise time that is short on the timescale of ferroelectric switching (for the
particular material in question). The initial response is therefore due to the non-ferroelectric polarization component,
which is quasistatic on the timescale of the input risetime.}
\\

\item{Initial Ferroelectric Response. A few $\mu$s after the initial pulse, the ferroelectric domains begin to switch.
This is the expected timecsale for the ferroelectric response of HfZrO materials \cite{Namlab}. As the domains begin to switch, the
capacitance of the Fe layer increases dramatically, increasing the charging current and resulting in a increased voltage
drop across the access resistor. As a consequence, the voltage across the capacitor stack drops.}
\\
\item{Final Ferroelectric Response. After the delayed rapid increase in ferroelectric capacitance, the ferroelectric capacitance gradually falls (the expected behavior, based on the P-V curve of Fig. \ref{PVModel}). The stack voltage continues to rise
(but on a slower timescale), eventually reaching VDD.}
\\
\end{enumerate}

The qualitative behavior of the ``spike" is captured well by the delay model. The final phase of the pulse response (phase 3) is captured only approximately; the empirical delay model indicates a ``standard" exponential-like RC response, whereas the data show a more gradual (almost linear) rise. This is not surprising, from the standpoint of ferroelectric delay. In the final stages of switching, only domains with a high switching threshold remain unswitched. These domains also have the slowest response. The empirical delay model does not treat the response of the various domains on an individual basis; instead, the switching dynamics are lumped into a single, unified response. This simplification makes it difficult to precisely capture the switching waveform. The basic separation of the response into the non-ferro initial response followed by the delayed ferro response is nevertheless clear.

The data (and simulation) also reveal an asymmetry in the switching response in the positive and negative direction. 
Whereas the positive ``spike" occurs with a stack voltage of approximately 4V, the negative one occurs much sooner, with a 
stack voltage of about -2V. This asymmetry in transient response is a consequence of the asymmetry of the coercive voltages
of the FeCap (which are approximately -1.5V and 3V, respectively). Since the peak of the ``spike" takes place shortly after 
the onset of ferroelectric switching, it is to be expected that the negative spike should occur much sooner than the positive one.  

\begin{figure}[!ht]
\centering
\includegraphics[width=3.6in]{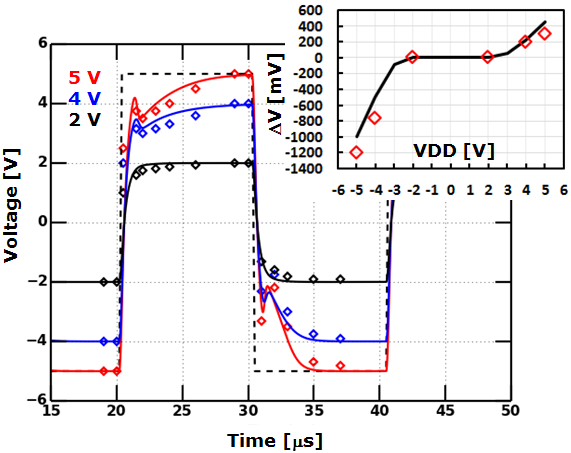}
% where an .eps filename suffix will be assumed under latex, 
% and a .pdf suffix will be assumed for pdflatex; or what has been declared
% via \DeclareGraphicsExtensions.
\caption{The measured \cite{SDattaEDL} and simulated transient responses for the ferroelectric capacitor stack
to bipolar square-pulses of various amplitudes are shown. The ``anomalous" spikes are diminishing in magnitude
with reducing input pulse amplitude (in both data and simulation). At a pulse voltage of 2V, the spikes are absent.
The inset shows the spike magnitudes ($\Delta$V) as a function of the applied voltage pulses, for both positive
and negative pulses. As in the main plot, the diamonds are from measured data \cite{SDattaEDL}, the lines
are from simulation.}
\label{Bipolar_VDD}
\end{figure}

\begin{figure}[!ht]
\centering
\includegraphics[width=3.5in]{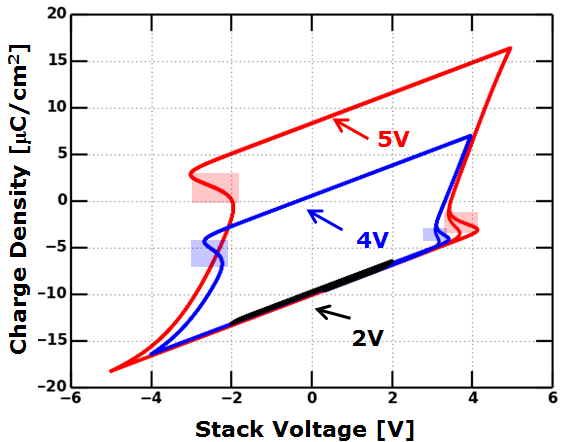}
% where an .eps filename suffix will be assumed under latex, 
% and a .pdf suffix will be assumed for pdflatex; or what has been declared
% via \DeclareGraphicsExtensions.
\caption{The simulated hysteresis for bipolar switching for several VDDs is shown. Regions of apparent negative
capacitance ($dQ/dV<0$) are shown shaded. At 5V and 4V, there are clear NC regions (though transient in nature),
while the 2V hysteresis loop is too tight to produce NC.}
\label{Bipolar_VDD_hyst}
\end{figure}

Additional insight into the switching behavior is gained by examining the VDD behavior of the switching. The same square-pulse waveform is applied, but with varying amplitudes. The data and simulation results are compared in Fig. \ref{Bipolar_VDD}, and the VF-Q switching characteristic is shown in Fig. \ref{Bipolar_VDD_hyst}. It is apparent that the "spike" effect is diminishing with decreasing pulse amplitude. While quite pronounced at 5V, it is completely gone at 2V. This diminishing trend is reproduced in simulation. Arguing using the ferroelectric delay model, the reason is simply that at low pulse voltages the voltage drop across the FeCap is too small to trigger ferroelectric domains. From Fig. \ref{PVModel}, it is evident that very little ferroelectric switching happens in the voltage range of [-1.5V, 1.5V], which is the approximate voltage range across the FeCap during the 2V pulse. In the Verilog-A model, the delay is associated only with ferroelectric switching, so absent switching, there is negligible delay (and hence no spike). An additional expected effect (though not modeled) is the increased switching time
for domains at low voltages \cite{Namlab}. This is currently being further investigated. At the $\mu$s timescale of the 
present data and simulations, the simple absence of domains in the small voltage range appears to be enough to explain the data. 

%\FloatBarrier
\subsection{Unipolar Switching}
\label{sec:Unipolar}

\begin{figure}[!ht]
\centering
\includegraphics[width=3.5in]{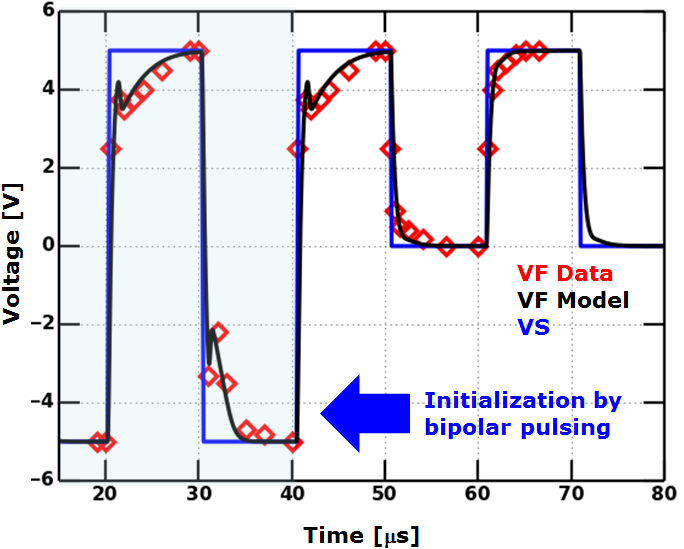}
% where an .eps filename suffix will be assumed under latex, 
% and a .pdf suffix will be assumed for pdflatex; or what has been declared
% via \DeclareGraphicsExtensions.
\caption{The measured \cite{SDattaEDL} and simulated transient response for the ferroelectric capacitor stack
to unipolar square-wave pulses is shown. The stack is initialized by a sequence of bipolar pulses (shaded blue region),
followed by a sequence of unipolar pulses. On the first unipolar pulse, the ``anomalous" spike is observed. On subsequent
pulses, the spike is absent.}
\label{Unipolar}
\end{figure}

\begin{figure}[!ht]
\centering
\includegraphics[width=3.4in]{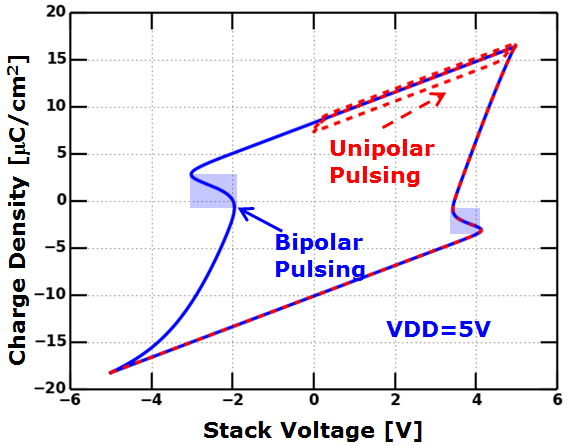}
% where an .eps filename suffix will be assumed under latex, 
% and a .pdf suffix will be assumed for pdflatex; or what has been declared
% via \DeclareGraphicsExtensions.
\caption{The P-V trajectory of the stack during bipolar and unipolar pulsing is shown. The initial pulses are
bipolar (blue solid line), establishing a wide loop with a large $\Delta$P. Subsequent pulses are unipolar
(red dashed line), resulting in a tight minor loop with minimal ferroelectric $\Delta$P (the initial portion of the unipolar
curve matches the bipolar since it switches from -5V to +5V). No apparent NC is exhibited on the minor loop.}
\label{Unipolar_hyst}
\end{figure}

For unipolar switching, the square input waveform is modified to range from 0 to VDD. The results are qualitatively different than
for the bipolar case. As can be seen in Fig. \ref{Unipolar}, the ``spike" is observed only on the first unipolar pulse of the waveform.
Subsequent pulses (in either the positive or negative direction) exhibit no spike (even at the highest VDD of 5V). This behavior
is reproduced in simulation, and is in fact largely independent of the details of the delay model. The reason for the behavior
of the stack voltage is apparent from Fig. \ref{Unipolar_hyst}. While the polarization undergoes large changes during bipolar
switching, the only large polarization change during unipolar switching is during the transition from bipolar to unipolar mode.
After this initial pulse, the FeCap operates on a tight minor loop, and the polarization changes are small and mostly
due to non-Ferro polarization.
As previously discussed, non-Ferro polarization is essentially quasistatic, and the ferroelectric $\Delta$P is very small.  
Together, the two effects cannot generate the transient spike. 
Modeling this behavior does not required a detailed delay model, just one
that delays only ferroelectric switching. Additionally, the transition to a tight minor loop must be captured, but this is 
reasonably straightforward with the Preisach model.

\FloatBarrier
\section{Impact on Subthreshold Slope}

The technological impetus for investigating the NC-effect is the potential improvement 
in FET subthreshold slope (SS). As argued in \cite{Datta}, a negative capacitance gate
layer can result in a sub-60 mV/dec SS, by introducing ``amplification" into the 
surface potential ($\psi$). Specifically, if the applied gate bias is V${_g}$, the 
long channel slope of the I$_d$-V$_g$ curve is proportional to $d\psi/d V_g$. With ordinary capacitors, this
derivative is never greater than unity (equal to unity only for long-channel, fully-depleted
devices). As shown in \cite{Datta}, NC effects can push the derivative beyond unity,
thereby reducing the SS below the usual theoretical limit. 
 \begin{figure}[!ht]
\centering
\includegraphics[width=3.4in]{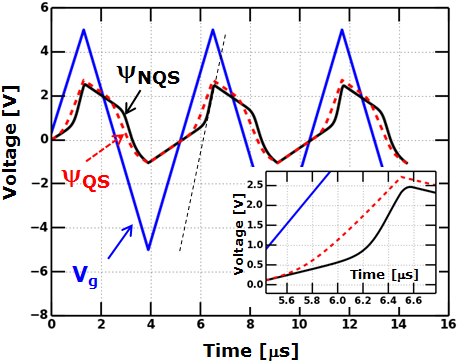}
% where an .eps filename suffix will be assumed under latex, 
% and a .pdf suffix will be assumed for pdflatex; or what has been declared
% via \DeclareGraphicsExtensions.
\caption{The surface potential $\psi$ and gate voltage V$_g$ are shown for a series of bipolar
switching events. In the context of Fig. \ref{Experiment}, the value of R is set to zero for this set
of simulations. As each switching event starts, $\psi$ initially tracks V$_g$ slowly, since the ferroelectric
domains are not yet switching. When the domains start to switch, $\psi$ changes rapidly. The approximate 
average slope
during the ferro switching interval is indicated by the black dashed line (and is seen to be somewhat higher than
the slope of V$_g$. The value of $\psi$ in the case of quasistatic switching ($\omega_0 \rightarrow \infty$)
is shown with a red dashed line for reference. No slope enhancement is observed. The inset figure provides
a close-up during ferro switching, also indicating the voltage range over which slope enhancement takes place.}
\label{slope1}
\end{figure}

 \begin{figure}[!ht]
\centering
\includegraphics[width=3.4in]{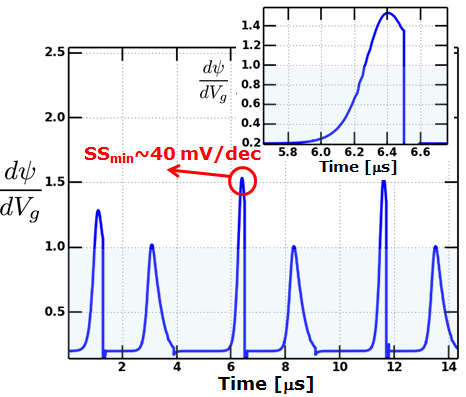}
% where an .eps filename suffix will be assumed under latex, 
% and a .pdf suffix will be assumed for pdflatex; or what has been declared
% via \DeclareGraphicsExtensions.
\caption{The slope $d\psi/dV_g$ is illustrated during bipolar switching events. It can be seen that
significant amplification occurs during brief periods, with  $d\psi/dV_g$ reaching 1.5 (corresponding
to a long-channel SS of 40 mV/dec). Sub-60 mV/dec slopes are attained for only ~0.5 $\mu$s, as can
be seen in the inset. }
\label{slope2}
\end{figure}

As can be seen in Figs. \ref{slope1} and \ref{slope2}, driving the capacitor stack with a frequency similar to
the ferroelectric switching frequency $\omega_0$ results in amplification over portions of the rising and
falling input voltage. The actual magnitudes of the slope $d\psi/dV_g$ are shown in Fig. \ref{slope2}; 
$d\psi/dV_g$ of approximately 1.5 is observed, resulting in a predicted SS of 40 mV/dec. It should be noted that
the improved SS is manifested only over small portions of the switching characteristic. Most of the simulated
points show un-amplified values of $d\psi/dV_g$. The latter are generally below unity because the underlying
semiconductor is not fully depleted (swinging from accumulation to inversion). Furthermore, before the ferroelectric domains start to switch, the
gate capacitance is small (only the non-Ferro component is present), resulting in unfavorable voltage division
between the (temporarily) small gate capacitance and the semiconductor (particularly true in regions where
the FET channel is in inversion). Consequently, $\psi$ is considerably
smaller than $V_g$ in this region, and so is  $d\psi/dV_g$. 

The simulation of  $d\psi/dV_g$ amplification was performed at VDD=5V; as seen in Fig. \ref{Bipolar_VDD},
voltages much below 5V do not result in the apparent NC effect for input pulse frequencies on the order of
a $\mu$s. It was argued that this is due to a lack of domain switching at low voltages. However, it was shown
in \cite{SDattaEDL} (and elsewhere) that low-voltage  $d\psi/dV_g$ amplification is indeed possible, but at much longer
timescales (approximately $10^4$ to $10^5$ times longer than simulated here). This may simply be a consequence 
of the much longer switching times of ferroelectric domains at low voltage (\cite{ferro2}, \cite{ferro3}, \cite{Namlab}).
The voltage dependence of the switching frequency characteristics is not modeled in the present work, and 
is the subject of current investigation.  

\FloatBarrier
\section{Conclusion and Future Direction}

A ferroelectric switching delay model is proposed to explain the ``anomalous" transient response
of ferroelectric capacitor stacks. While the ``anomalous" spike observed in the transient response has
been attributed to the NC effect, it is argued herein that this is not in fact necessary. By constructing an 
FeCap model which consists of a delayed and quasistatic component (the ferroelectric and electronic
polarization components, respectively), it is possible to reproduce the measured data in simulation
with reasonable fidelity.
This suggests that the observed NC effect in the data may simply be a dynamic (non-quasi static)
effect of the natural ferroelectric delay. Furthermore, there is clear evidence that the NC ``spike"
is tied directly to ferroelectric switching:\\
\begin{itemize}
\item The onset of the end of the ``spike" is correlated to the ferroelectric coercive voltage (as evidenced
in Sec. \ref{sec:Bipolar}).\\
\item The magnitude of the ``spike" (or its presence) is correlated to the extent to which ferroelectric domains
switch (as evidenced in Sec. \ref{sec:Bipolar}, Fig. \ref{Bipolar_VDD}).\\
\item The presence of the ``spike" is correlated to conditions in which large changes in ferroelectric polarization
are observed, i.e. bipolar and not unipolar switching (as evidenced in Sec. \ref{sec:Unipolar}). Furthermore, the
presence or absence of the ``spike" correlates with the polarization history.\\
\end{itemize}

While the ferroelectric delay model does successfully reproduce the transient pulse data, it must be recognized
that there are numerous other phenomena in the literature associated with NC.  Additional experimental and modeling
work is required to fully explain all observations. Of specific interest is the apparent voltage amplification and the associated
sub-60 mV/dec SS behavior of FeFETs with capacitor stacks such as those explored in this paper. While the delay model
predicts amplification and sub-60 mV/dec behavior for the stack in question, it suggests that it cannot occur on the 
timescale of the experimental data and technologically desired low voltages (VDD $\approx$ 1V). 
This is obvious from the VDD behavior of the
pulse data in Sec. \ref{sec:Bipolar}. However, amplification may well occur even at low VDD on a much slower timescale,
given the much slower ferroelectric switching at low voltages. It is therefore desirable to further investigate experimentally
the long-timescale, low-voltage regime. The ferroelectric delay model will likewise need to improved to capture the effects
of slow switching at low voltage.
  
From a scaled technology perspective, an individual device and inverter switches on a few-ps timescale. 
This limits the usefulness of sub-60 mV/dec transistors as a general purpose scaled device, if indeed these devices switch at a much slower timescale. FeFETs with moderately high RO frequencies have already been reported \cite{GF}, but the reduced RO frequency relative
to standard dielectrics suggests than only non-Ferro switching is taking place (consistent with a decrease in power, in spite of
the obviously high DC gate capacitance). 
In order to properly explore the frequency dependence, a pulse train similar to \cite{Namlab} needs to be applied to the gate terminal of the FeFET transistor with varying voltage magnitude and pulse duration both in forward and reverse direction. This set of measurements is the key to decipher the underlying mechanism, whether it is truly NQS or not, and uncover the technology implications for FeFETs. Barring the aforementioned measurements, a technological evaluation of FeFETs will remain incomplete. Evolution of the sub-60 behavior as a function of temperature may shed light on the underlying physics as well. Another point worth noting is that the range of Vg for which the SS is less than 60 mV/dec is a critical parameter to track for possible technology implications. If indeed the Vg range is very small for a SS reduction to say 40 mV/dec, significantly lower leakage or lower Vt may not be realized for the entire voltage range of technological interest, limiting applicability of these devices even for a slower applications.

%\begin{figure}[!h]
%\centering
%\includegraphics[width=2.5in]{LongChannelIMEC.png}
%% where an .eps filename suffix will be assumed under latex, 
%% and a .pdf suffix will be assumed for pdflatex; or what has been declared
%% via \DeclareGraphicsExtensions.
%\caption{The room-temperature Id-Vg characteristics of the long-channel In$_{70}$Ga$_{30}$As GAA are illustrated.
%PBE is negligible at this length. The leakage floor is set by BTBT alone.  Symbols are measured data, lines
%are simulation results.}
%\label{LongChannelIMEC}
%\end{figure}

% or


\section*{Acknowledgment}

The authors would like to kindly thank Prof. Suman Datta and his group at the 
University of Notre Dame for very helpful discussions.

\ifCLASSOPTIONcaptionsoff
  \newpage
\fi



%\bibliographystyle{IEEEtran}

\begin{thebibliography}{1}


\bibitem{Datta}
S.~Salahuddin, S.~Datta
"Use of Negative Capacitance to Provide
Voltage Amplification for Low Power
Nanoscale Devices"
Nano Letters 2008 Vol. 8, No. 2.

% Time-dependent positive
% capacitance model according to Komogolov-Avrami-Ishibashi theory (red line) and time-dependent Ri model
%(blue line) using in PSPICE simulator


\bibitem{Sayeef}
A. I.~Khan, K.~Chatterjee, B.~Wang, S.~Drapcho, L.~You, C.~Serrao,
S. R.~Bakaul, R.~Ramesh, S.~Salahuddin
``Negative capacitance in a ferroelectric capacitor"
Nature Materials Letters, December 15th 2014.


\bibitem{ferro1}
J.~Muller, , T.S.~Boscke, D.~Brauhaus, U.~Schroder, U.~Bottger, J.~Sundqvist, P.~Kücher, T.~Mikolajick, L.~Frey 
``Ferroelectric Zr0.5Hf0.5O2 thin films for nonvolatile memory applications",
Appl. Phys. Lett. 99, 112901 (2011).


\bibitem{SDattaEDL}
P.~Sharma, J.~Zhang, K.~Ni, S.~Datta
``Time-Resolved Observation of Negative Capacitance"
IEEE Electron Device Letters 2018.

\bibitem{Stabilization}
A.~Jain, M. A.~Alam, 
``Stability Constraints Define the Minimum Subthreshold Swing of a 
Negative Capacitance Field Effect Transistor" IEEE Trans. Elec. Dev. 61 2235 (2014). 

\bibitem{Preisach}
G.~Robert, D.~Damjanovic, N.~Setter
``Preisach modeling of ferroelectric pinched loops"
Appl. Phys. Lett. 77, 4413 (2000).

\bibitem{Preisach2}
A. T.~Bartic, D. J.~Wouters, H. E.~Maes, J. T. ~Rickes, and R. M.~Waser
``Preisach model for the simulation of ferroelectric capacitors"
Journal of Applied Physics 89, 3420 (2001).

\bibitem{GF}
Z.~Krivokapic, U.~ Rana, R.~Galatage, A.~Razavieh, A.~Aziz, J.~Liu, J.~Shi, H.J.~Kim, R.~Sporer, C.~Serrao,
A.~Busquet, P.~Polakowski, J.~Müller, W.~Kleemeier, A.~Jacob, D.~Brown, A.~Knorr, R.~Carter, S.~Banna
``14nm Ferroelectric FinFET Technology with Steep
Subthreshold Slope for Ultra Low Power Applications"
IEEE IEDM 2017.

\bibitem{alternative}
S. J.~Song, Y. J.~Kim, M. H.~Park, Y. H.~Lee, H. J.~Kim,
T.~Moon, K. D.~Kim, J.-H.~Choi, Z.~Chen, A.~Jiang 
C. S. Hwang
``Alternative interpretations for decreasing voltage with increasing
charge in ferroelectric capacitors"
Nature Scientific Reports, February 2016.


\bibitem{ferro2}
J.~Li, B.~Nagaraj, H.~Liang, W.~Cao, C.~H.~Lee, R.~Ramesh,
"Ultrafast polarization switching in thin-film ferroelectrics"
Appl. Phys. Lett. 84, 1174 (2004).

\bibitem{ferro3}
M.~Kobayashi, N.~Ueyama, K.~Jang, T.~Hiramoto,
``Experimental Study on Polarization-Limited Operation Speed of
Negative Capacitance FET with Ferroelectric HfO2",
IEEE International Electron Devices Meeting (IEDM), 2016.

\bibitem{Namlab}
H.~Mulaosmanovic, J.~Ocker, S.~Müller, U.~Schroeder, J.~Müller,
P.~Polakowski, S.~Flachowsky, R. v.~Bentum, T.~Mikolajick, S.~Slesazeck
``Switching Kinetics in Nanoscale Hafnium Oxide Based Ferroelectric
Field-Effect Transistors"
ACS Applied Materials and Interfaces, 2017, 9, 3792 −3798.


\end{thebibliography}
\FloatBarrier

\end{document}